\documentclass{article}

\usepackage{graphicx}
\usepackage{cite}

\usepackage{color}

\def\giorno{9/2/2020}

\def\a{\alpha}
\def\b{\beta}
\def\ga{\gamma}
\def\vphi{\varphi}

\def\s{\sigma}

\def\vphi{\varphi}

\def\F{{\mathcal F}}
\def\G{{\mathcal G}}

\def\pa{\partial}

\def\o+{\oplus}

\def\Lap{\triangle}  %% laplacian

\def\<{\langle}
\def\>{\rangle}

\def\({\left(}
\def\){\right)}
\def\[{\left[}
\def\]{\right]}
\def\=#1{\bar #1}
\def\~#1{\widetilde #1}

\def\.#1{\dot #1}
\def\^#1{\widehat #1}

\def\"#1{\ddot #1}

\def\eeq{\end{equation}}
\def\beq{\begin{equation}}

\def\beql#1{\begin{equation} \label{#1}}

\def\eqref#1{(\ref{#1})}

\def\EOR{ \hfill $\odot$ \medskip}

\def\EOP{ \hfill $\triangle$ \medskip}

\def\riq#1{#1}

\def\symmref{AVL,CGbook,KrV,Olver1,Olver2,Stephani}
\def\sderef{Arnold,Evans,Fre,Ikeda,Kampen,Oksendal,Stroock}
\def\stochsymmref{GRQ1,GRQ2,Unal,Koz1,Koz2,Koz3,GS17,GGPR,GL1,GL2,Koz18a,Koz18b,KozB,GLS,GW18}

\begin{document}

\title{Symmetry classification of scalar Ito equations with multiplicative noise}

\author{Giuseppe Gaeta\thanks{
Dipartimento di Matematica, Universit\`a degli Studi di Milano, via Saldini 20133 Milano (Italy)
{\tt and}
SMRI, 00058 Santa Marinella (Italy);
e-mail: {\tt giuseppe.gaeta@unimi.it}}${\ }^\ddag$ \ \& Francesco Spadaro\thanks{EPFL-SB-MATHAA-CSFT, Batiment MA - Station 8,
CH-1015 Lausanne (Switzerland); e-mail: {\tt francesco.spadaro@epfl.ch}}{\ }\thanks{GNFM-INdAM}}

\bigskip

\date{\giorno}

\maketitle

\begin{abstract}

\noindent We provide a symmetry classification of scalar
stochastic equations with multiplicative noise. These equations
can be integrated by means of the Kozlov procedure, by passing to
symmetry adapted variables.

\end{abstract}

\section{Introduction}
\label{sec:intro}

In a recent paper \cite{G19} we have explicitly integrated the
\emph{stochastic logistic equation with multiplicative noise}
\beql{eq:SLE} d x \ = \ \( A \, x \ - \ B \, x^2 \) \ d t \ + \
\mu \, x \ d w \ , \eeq where  $A,B,\mu$  are positive real
constants. The motivation for this was provided by questions in
Mathematical Biology, in particular Population Dynamics,  where
the logistic equation plays a central role. In this context,
multiplicative noise is also known as \emph{environmental noise},
as it models fluctuations due to changes in the environmental
conditions and thus acting in the same way -- and in a fully
correlated manner -- on all the individuals
\cite{OvaMee,NoiseRev}. (Uncorrelated fluctuations for different
individuals give raise to noise terms proportional to $\sqrt{x}$;
in that context, one refers to this as \emph{demographic noise}
\cite{OvaMee}.)

In order to integrate \eqref{eq:SLE}, we have employed tools from
the recently developed theory of \emph{symmetry of stochastic
differential equations} \cite{\stochsymmref}. In particular, it
was found that \eqref{eq:SLE} admits a simple Lie symmetry and
hence, thanks to a general constructive theorem by Kozlov
\cite{Koz1,Koz2,Koz3} (see also \cite{GGPR,GL1,GL2,GLS}), one can
pass to symmetry-adapted variables allowing for a direct
integration. See \cite{G19} for details.

It is natural to wonder if the integrability of \eqref{eq:SLE} was
a lucky accident or, as it indeed appears more probable,
\eqref{eq:SLE} is one of a more and less wide family of integrable
equations.

In the present work, we want to consider scalar Ito equations (we
always assume the drift $f(x,t)$ and noise $\s (x,t)$ are smooth
functions of their arguments) \beql{eq:Ito} d x \ = \ f(x,t) \, d
t \ + \ \s (x,t) \, d w \eeq with a noise term $\s (x,t)$
corresponding to multiplicative noise, i.e. $$ \s (x,t) \ = \ S(t)
\ x \ ; $$ we assume $S(t) \not\equiv 0$, or we would have a
deterministic equation. For equations in this class, we want to
classify those admitting a simple Lie-point symmetry and which
hence can be integrated by means of the Kozlov approach.

Thus the general form of equations to be considered is
\beql{eq:eq} \riq{ d x \ = \ f(x,t) \ d t \ + \ S(t) \, x \ d w }
\ . \eeq We will look for the most general symmetries allowed by
our classification \cite{GS17}, i.e. W-symmetries; we will however
restrict our attention to \emph{simple} W-symmetries, as these are
the only ones which can be used to integrate the equation
\cite{GW18}. That is, we look for symmetry vector fields of the
form \beql{eq:X} X \ = \ \vphi (x,t,w) \, \pa_x \ + \ R \, w \,
\pa_w \ , \eeq
where $R$ is a real constant.

In this case the determining equations read \cite{GW18}
\begin{eqnarray}
\vphi_t &+& f \, \vphi_x \ - \ \vphi \, f_x \ + \ \frac12 \Lap (\vphi ) \ = \ 0 \label{eq:deq1} \\
\vphi_w &+& \s \, \vphi_x \ - \ \vphi \, \s_x \ = \ R \, \s  \ .
\label{eq:deq2} \end{eqnarray} Here and below $\Lap$ is the Ito
Laplacian, which in our scalar case reads simply \beq \Lap ( \Psi
) \ := \ \frac{\pa^2 \Psi}{\pa w^2} \ + \ 2 \, \s \, \frac{\pa^2
\Psi}{\pa x \pa w} \ + \ \s^2 \, \frac{\pa^2 \Psi}{\pa x^2} \ .
\eeq

In our discussion, we will assume the reader to be familiar with the standard theory of symmetry for deterministic differential equations \cite{\symmref} and 
with stochastic differential equations \cite{\sderef}. 

\medskip\noindent
{\bf Remark 1.} We recall that, as shown by Kozlov
\cite{Koz1,Koz2,Koz3} (see also \cite{GS17,GW18} for the extension
of Kozlov theorem to W-symmetries), once we have determined a
simple symmetry of an Ito SDE, this is explicitly integrated by
passing to the (symmetry adapted) new variable \beq y \ = \ \int
\frac{1}{\vphi (x,t,w)} \ d x \ ; \eeq thus our classification of
equations (with multiplicative noise) admitting a simple symmetry
also provides a classification of equations (in this class) which
can be explicitly integrated by the Kozlov change of variables.
\EOR

\medskip\noindent
{\bf Remark 2.} The paper \cite{Koz2} also provides a symmetry
classification of scalar SDEs; this refers to \emph{standard
deterministic} symmetries, while the present one deals with the
(more general) framework of \emph{W-symmetries}, including also
the case of \emph{standard random} symmetries \cite{GS17}. On the
other hand, in \cite{Koz2} one considers general scalar Ito
equations, while we only deal with a specific form of the
diffusion coefficient $\s (x,t)$. \EOR

\section{The basic classification}
\label{sec:classif}

The determining equations for our specific form of Ito equation
with multiplicative noise \eqref{eq:eq} restrict the scenario --
in terms of possible drifts term -- for which symmetries are
present: only when the drift $f(x,t)$ has a specific form, the
deterministic equations allow solutions $\varphi$, i.e.
symmetries (standard or W). At this point we do not want to
discuss about regularity requirements of the functions involved,
but we will rather derive the constrained form of $f(x,t)$. Thus,
from now on, we will assume all functions we meet to be regular
enough. In the following computation we will get the
aforementioned constraints on $f(x,t)$, summarized in Proposition
1, by solving the determining equations by separation of
variables.

For the specific form of \eqref{eq:eq}, the second determining
equation \eqref{eq:deq2} reads \beq \vphi_w \ + \ S \ \( x \,
\vphi_x \ - \ \vphi \ - \ R \, x \)  \ = \ 0 \ . \eeq The general
solution of this is\footnote{Note that by $\log (x)$ we will
always mean the natural logarithm; moreover it will be understood
that we write $\log (x)$ for $\log ( |x|)$. This is specially
justified when we look at equations like \eqref{eq:SLE}, with $x$
describing a population -- which is by definition non-negative --
which is the motivation of our work.} \beql{eq:phi} \riq{\vphi
(x,t,w) \ = \ x \
\[ R \, \log (x) \ + \ q(t,z) \] } \ ,  \eeq where we have written
\beql{eq:z} z \ := \ w \ - \ \frac{\log (x)}{S(t)}  \ . \eeq

Restricting \eqref{eq:deq1} to functions of this type, i.e.
plugging \eqref{eq:phi} into \eqref{eq:deq1}, we get an equation
of the form \beql{eq:deq1s} x \, q_t \ + \ \chi_0 \ + \ \chi_1 \,
q \ + \ \chi_2 \, q_z \ = \ 0 \ , \eeq where we have written
\begin{eqnarray*}
\chi_0 &=& \frac{R}{2} \ \[ 2 \, f  \, [1 + \log (x) ] \ + \ x \, \( S^2 \, - \, 2 \, \log (x) \, f_x \) \] \ , \\
\chi_1 &=& \( f \ - \ x \, f_x \) \ , \\
\chi_2 &=& \frac12 \, x \ S \ + \ \frac{x \, S'}{S^2} \, \log (x) \ - \ \frac{f}{S} \ . \end{eqnarray*}
Here both the functions $f(x,t)$ and $q (t,z)$ are unknown. Moreover, now our set of variables is $(x,t,z)$.

Differentiating \eqref{eq:deq1s} three times in $x$ and once in $z$, eliminating a common factor $(x^2 S(t))^{-1}$ (as stressed above $S(t) \not= 0$), and writing
\beql{eq:gf} g(x,t) \ := \ f_{xxx} (x,t) \eeq for ease of notation, we get
\begin{eqnarray} & &  q_{zz} (t,z) \ \[ S' (t) \ + \ x^2 \, S(t) \, g (x,t) \] \ + \nonumber \\
&  & \ + \ q_z (t,z) \ \[ 2 \, x^2 \, S^2 (t) \, g (x,t) \ + \ x^3
\, S^2 (t) \, g_x (x,t) \] \ = \ 0 \ . \label{eq:eqr0}
\end{eqnarray} Now we can separate the $x$ and $z$ variables,
which yields \beql{eq:eqr} - \ \frac{q_{zz}}{q_z} \ = \ \frac{2 \,
x^2 \, S^2 (t) \, g (x,t) \ + \ x^3 \, S^2 (t) \, g_x (x,t)}{S'
(t) \ + \ x^2 \, S(t) \, g (x,t)} \ . \eeq Here the l.h.s. is a
function of $z$ and $t$, while the r.h.s. is a function of $x$ and
$t$. Thus this relation can hold only if both sides are a function
of $t$ alone, i.e. if the l.h.s. is independent of $z$ and the
r.h.s. is independent of $x$.

Requiring that the l.h.s. of \eqref{eq:eqr} is independent of $z$ amounts to solving
\beq q_{zz}^2 \ = \ q_z \ q_{zzz} \ ; \eeq
this yields immediately (here $Q_i (t)$ are arbitrary smooth functions)
\beql{eq:qgen} \riq{ q(t,z) \ = \ \exp[z \, Q_1(t)] \ Q_2(t) \ + \ Q_3 (t) } \ . \eeq

\medskip\noindent
{\bf Remark 3.} Note that we can exclude the case $Q_1 (t)=0$. In
fact, if this was the case we would have $q(t,z) = Q_2 (t) +
Q_3(t)$, but as these are both arbitrary functions, it is the same
as assuming $q(t,z) = Q_3 (t)$, which is obtained for $Q_2 (t) =
0$. \EOR
\bigskip

As for the request that the r.h.s. of \eqref{eq:eqr} is
independent of $x$, this leads to \beql{eq:eqfg} 4 \, S' \, g \ +
\ x^3 \, S \, g \, g_x \ + \ 5 \, x \, S' \, g_x \ - \ x^4 \, S \,
g_x^2 \ + \ x^4 \, S \, g \, g_{xx} \ + \ x^2 \, S' \, g_{xx} \ =
\ 0 \ . \eeq
This equation can be solved, yielding the general
solution \beql{eq:ggen} g(x,t) \ = \ \frac{1}{x^2 \, S(t)} \ \[
x^{\a (t)} \  \eta(t)  \ - \ S'(t) \] \ , \eeq where again $\a$
and $\eta$ are arbitrary smooth functions.

\medskip\noindent
{\bf Remark 4.} We pause a moment to note that in order to
determine $f(x,t)$ from a $g(x,t)$ of this general form, we should
solve \eqref{eq:gf} as an equation for $f$. In doing this, we
should pay attention to the arbitrary function $\a(t)$: it turns
out that if $\a (t)$ is actually constant and takes either of the
values $\{ -1 , 0 ,+1\}$ we  have special cases (one of the three
integrations in $x$ concerns a factor $1/x$ and hence produces a
logarithm), as discussed below. \EOR
\bigskip

We can now go back to   \eqref{eq:eqr}; with the expression
obtained above for $q(t,z)$ and $g(x,t)$, this reads \beql{eq:Q1}
\riq{Q_1 (t) \ = \ - \ \a (t) \ S(t)}  \ . \eeq

Thus we set $Q_1 (t)$ to be as in \eqref{eq:Q1}, and with this we
get \beq q(t,z) \ = \ \exp \[ - \, \a(t) \, S(t) \, z \] \ Q_2 (t)
\ + \ Q_3 (t) \ . \eeq

\medskip\noindent
{\bf Remark 5.} It may be noted that in this way we constrain the
expression for (the possible) $\vphi$; in particular, we have
\beql{eq:phigen} \riq{ \vphi (x,t) \ = \ x \ \[ R \, \log(x) \ + \ x^{\a (t)} \
\exp [- \a (t) S(t) w] \ Q_2 (t) \ + \ Q_3 (t) \] } \ . \eeq Thus
$\vphi$ can indeed depend on $w$ (unless $\a(t)$ or $Q_2 (t)$ are
zero). Note also that (at least at this stage) we can have proper
W-symmetries, i.e. $R \not= 0$; moreover the possible dependence
on $\log (x)$ is related to $R \not= 0$ -- thus  standard
symmetries will not have any dependence on $\log (x)$. \EOR
\bigskip

Let us now go back to considering $f(x,t)$, i.e. the possible form
of the symmetric Ito equations. Now that we have an expression for
$g(x,t)$ the form of $f(x,t)$ is obtained by integrating it, i.e.
solving \eqref{eq:gf} as an equation for $f$.

For the generic case, which in this context means $\a(t) \not=
-1,0,+1$ (these cases should be dealt with separately), this
yields \beql{eq:fgen} \riq{f(x,t) \ = \ \[ F_1 (t) \ + \ x \ F_2 (t) \
+ \ x^2 \ F_3 (t) \] \ + \ \Psi (x,t) } \eeq where we
have defined \beql{eq:Psigen} \Psi (x,t) \ := \  \[ \( \log (x) \, - \,
1 \) \, x \ \frac{S'(t)}{S(t)} \ - \ \Phi (x,t) \ \frac{\eta(t)}{S(t)} \]  \ , \eeq
\beql{eq:Phigen} \Phi (x,t) \ := \ \,  \ \frac{x^{1+\a (t)} }{[1
-\a(t)] \, \a (t) \, [1 + \a (t) ]} \ . \eeq

As mentioned above, this expression holds provided $\a (t)$ is not
constant and equal to either $0$ or $\pm 1$. In these special
cases, the expression \eqref{eq:fgen} still holds\footnote{Indeed
the term in square brackets in the r.h.s. of \eqref{eq:fgen}
correspond to the constants of integration -- which of course
depend on $t$ -- in the three $x$ integrations.}, but now $\Phi
(x,t)$ should be defined in a different way:
\begin{eqnarray}
\a(t) \ = \ - 1 & \ \ & \Phi (x,t) \to \Phi_{(-)} (x,t) \ := \ - \ \frac12 \ \log (x)  \ , \\
\a(t) \ = \ \ 0 & \ \ & \Phi (x,t) \to \Phi_{(0)} (x,t) \ := \ - \ x \ \[ 1 \ - \ \log(x)\] \ , \\
\a(t) \ = \ + 1 & \ \ & \Phi (x,t) \to \Phi_{(+)} (x,t) \ := \ - \
\frac14 \ x^2 \ \[ 2 \log(x) \ - \ 3 \] \ . \end{eqnarray}

We can insert these expressions in that for $f(x,t)$; one should recall here that the $F_k (t)$ are generic functions, so that adding to them any other function of time will not change their nature but just change them from $F_k (t)$ to some other function $\^F_k (t)$, which we will still write as $F_k (t)$.

In this way, and writing for ease of notation
$$ \Sigma (t) \ := \ \frac{S'(t)}{S(t)} \ , \ \ \mathcal{F} (x,t) \ = \ F_1 (t) \ + \ x \ F_2 (t) \
+ \ x^2 \ F_3 (t) \ , $$  we have
\beq f(x,t) \ = \ \cases{\mathcal{F} (x,t) \ + \ G(t) \, x^{1 +\a(t)} \ + \ \Sigma(t) \, x \, \log x & ($\a (t)$ generic) \cr
\mathcal{F} (x,t) \ + \ \Sigma (t) \, x \, \log x \ + \ H(t) \, x^2 \, \log x & ($\a(t) = +1$)  \cr
\mathcal{F} (x,t) \ + \ H(t) \, x \log x & ($\a (t) =0 $) \cr
\[ F_1 (t) \ + \ x \ F_2 (t) \
+ \ x^2 \ F_3 (t) \] \ + \ H(t) \, \log x & ($\a (t) = - 1$) \cr} \eeq
Here $G(t)$ and $H(t)$ are arbitrary functions.

Before going on to discuss the different cases, we note that our discussion so far already identifies the -- rather restricted -- general class of equations of the form \eqref{eq:eq} which \emph{could} admit some symmetry. We summarize the result of our discussion in this respect in the following Proposition 1; note that in there we have used the presence of arbitrary functions, and also write
$$ \mathcal{H} (x,t) \ = \ H_1 (t) \ + \ x \ H_2 (t) \
+ \ x^2 \ H_3 (t) \ , $$
in order to further simplify the writing of our functional forms.

\medskip\noindent
{\bf Proposition 1.} {\it Scalar SDEs of the form \eqref{eq:eq} may admit a symmetry only if $f(x,t)$ is of the general form
\beql{eq:prop1} f(x,t) \ = \  \mathcal{F} (x,t)  \ + \ G (t) \, x^{1 + \a (t)} \ + \ \mathcal{H} (x,t)  \ \log x \ .  \eeq}

\medskip\noindent
{\bf Remark 6.} It may be worth mentioning that this is a general form and provides necessary, but by no means sufficient, conditions for the existence of a symmetry; under closer scrutiny we will find out that actually this is too general to provide also sufficient conditions: e.g., a symmetry will be present only if $H_3 (t)= 0$. \EOR

Let us come back to our task of classifying symmetric equations and symmetries. We should now insert these expressions for $q(t,z)$ and $f(x,t)$
in \eqref{eq:deq1}. We obtain quite complicate expressions --
which we do not report here -- in the generic and in the three
special cases.

The relevant point is that all dependencies on $x$ -- including
those in $\log (x)$ -- and $z$ are now explicit, so that
coefficients of different monomials in $\{z,x,\log(x)\}$ should
vanish separately, and the equation splits in several simpler
ones.

\medskip\noindent
{\bf Remark 7.} It should also be noted that in our equations we
have some factors $x^{\a (t)}$, see \eqref{eq:ggen}; if $\a (t)$
is actually a constant and in particular if $\a (t)$ is an
integer, this can interact with other terms. This can actually
happen only for $\a(t) = 0,\pm 1$, see \eqref{eq:prop1}, and we have already warned that these should be treated separately (exactly for this reason); we
will thus refer to the case where $\a (t) \not= 0,\pm 1$ as the
\emph{generic} case, and to those where $\a(t) = 0,\pm 1$ as
\emph{special} cases. \EOR

\medskip\noindent
{\bf Remark 8.} Note also that in the special cases we can always
set $\eta (t) = 0$ with no loss of generality. \EOR

\section{Results}

We will give here our results; the proof, which consists largely
of a detailed computation, is confined to the ``supplementary
material'' (see below).

In order to state our results more compactly, it will be
convenient to introduce the notations \beql{eq:Gamma} \Gamma_\pm
(t) \ := \ \frac{S^2(t)}{2} \ \pm \ \frac{S(t)}{ k} \ \[ \Sigma
(t) \ - \ \chi (t) \] \ ; \ \ \ \chi (t) \ := \ \frac{Q'(t)}{Q(t)} \
. \eeq

\medskip\noindent
{\bf Theorem 1.} {\it The equations of the form \eqref{eq:eq}
admit standard ($R=0$) or W-symmetries ($R \not= 0$) for $f(x,t)$
given by one of the expressions in the following table, where $\F$
and $\G$ and $\theta$ are arbitrary functions (possibly zero):
$$ \begin{tabular}{||l||c|c||l||}
\hline
 case & $S(t)$ & $R$ & $f(x,t)$ \\
\hline
\hline
(a) & $S(t)$ & 0 & $\Gamma_+ (t) \, x \ + \ \Sigma(t) \, x \, \log x \ + \ \G(t) \, x^{1 + k/S(t)}$ \\
(b) & $S(t)$ & 0 & $\Gamma_+ (t) \, x \ + \ \Sigma(t) \, x \, \log x$ \\
(c) & $S(t)$ & 0 & $\F (t) \, x \ + \ \Sigma(t) \, x \, \log x$  \\
(d) & $S(t)$ & 0 & $\F (t) \, x \ + \ \theta(t) \, x \, \log x$ \\
(e) & $S(t)$ & $R$ & $- \Gamma_- (t) \, x \ + \ \Sigma (t) \, x \, \log x$ \\
 \hline
(f) & $s_{0}$ & 0 & $\F (t) \, x^2 \ + \ [ s_0^2/2 - \chi (t) ] \, x$ \\
(g) & $s_0$ & 0 & $\F (t) \ + \ [ s_0^2/2 - \chi (t) ] \, x$ \\
\hline
(h) & $s_0$ & $R$ & $[s_0^2/2 - \chi (t)] \, x$ \\
 \hline
\end{tabular} $$

Their (standard or W) symmetries are given by \eqref{eq:X} with $R$ as in the previous table and coefficient $\vphi$ given by the following table, where $k \not=0$ and $K$ (possibly $K=0$) are arbitrary constants:
$$ \begin{tabular}{||l||c|l||}
\hline
 case & $R$ & $\vphi (x,t,w)$ \\
 \hline
 \hline
(a) & 0 & $Q(t) \, e^{-k w} \, x^{1 +k/S(t)}$ \\
(b) & 0 & $K \, S(t) \ + \ Q(t) \, e^{-k w} \, x^{1 +k/S(t)}$\\
(c) & 0 & $K \, S(t) \, x $ \\
(d) & 0 & $\theta(t) \ x $ \\
(e) & $R$ & $R \[ \F (t) \, e^{-k w} x^{1 +k/S(t)} \ + \ \theta (t) \, x \ + \  x \log x \] $ \\
 \hline
(f) & 0 & $K \,s_0 \, x \ + \ Q(t) \, e^{- s_0 w} \, x^2 $ \\
(g) & 0 & $K \,x \ + \ Q(t) \, e^{s_0 w} $ \\
 \hline
(h) & $R$ & $R [ \theta (t) x + J(t) e^{ s_0 w} x^2 + x \log x]$ \\
 \hline
\end{tabular} $$

\medskip\noindent
Furthermore, no other scalar Ito stochastic differential equations
\eqref{eq:eq} beside those detailed here admit standard or
W-symmetries.}

\medskip\noindent
{\bf Proof.} As mentioned before, the proof consists in explicitly
computing and carrying out, starting from Proposition 1 and the
determining equations, the possible symmetries listed in the table
above. Checking that these equations admit the given symmetries
just requires direct elementary computations, i.e. checking that
the determining equations \eqref{eq:deq1}, \eqref{eq:deq2} are
satisfied. The second part of the statement, i.e. that these are
the only equations admitting symmetries, requires more involved
computations. For a sake of readability, these computations are
given explicitly in the supplementary material. \EOP

\section{Discussion and Conclusions}

We have considered in full generality equations of the form
$$ d x \ = \ f(x,t) \ d t \ + \ S(t) \ x \ d w \ , $$
i.e. scalar stochastic differential equations with multiplicative
noise, also known in population dynamics as equations with
environmental noise.

We have investigated in which cases, i.e. for which functions
$f(x,t)$ and $S(t)$, they admit a symmetry (or more precisely a
one-parameter group of symmetries), setting this search within the
most general class of symmetries, i.e. W-symmetries.

Our detailed analysis had to consider a generic case and three
special cases, depending on the function $\a (t)$ introduced in
\eqref{eq:ggen}; the special cases correspond to $\a (t)$ being a
constant function with value $\{-1,0,+1\}$.

We have found that the most general functional form of $f(x,t)$
admitting a symmetry (including the special cases) is
\beql{eq:ffin} \riq{ f(x,t) \ = \ \mathcal{F} (t) \ x \ + \
\mathcal{G} (t) \ x^{1 + \a (t)} \ + \ \mathcal{H} (t) \ x \ \log
(x) } \ . \eeq

As for the symmetries, these are characterized by $R$ and $\phi$,
see \eqref{eq:X}; the most general form of $\vphi$ turned out to
be \beql{eq:phifin} \riq{ \vphi(x,t,w) \ = \ \mathcal{A} (t) \ x \
+ \ \mathcal{B} (t) \ \exp \[ k \, w \] \ x^{1 + \a(t)} \ + \ R \
x \ \log x } \ . \eeq

The reader is referred to Theorem 1 for the specific form taken by
the functions $\{ \mathcal{F} (t), \mathcal{G} (t) , \mathcal{H}
(t); \mathcal{A} (t) , \mathcal{B} (t) \}$ as well as for $\a (t)$
and $S(t)$, which are in many cases constrained; in particular,
several of the specific cases obtained above require that $S(t)$
is constant. Note also that in all subcases admitting such terms
but one, we had $\mathcal{H} (t) = [S' (t)/S(t)]$; this also means
that the terms $x \log x$ can be allowed in $f(x,t)$ only if
$S(t)$ is actually constant.

It should be stressed that our work points out at a substantial
difference with respect to the deterministic case: in fact, in
that case the functional form of deterministic equations admitting
symmetries (and of the symmetries) is largely unconstrained, and
actually for the closet analogue to the equations considered here,
i.e. for dynamical systems, it is known that we always have
symmetries (albeit these can not be determined algorithmically).
We have found that the situation is completely different for
stochastic equations: not only symmetry is a non-generic property,
but the functional form of stochastic equations admitting
symmetries is severely constrained. This is maybe the most
relevant -- and qualitative -- result of our classification;
unfortunately it also means that, at least for scalar equations,
the value of symmetry methods is restrained to a rather specific
class of equations.

Our classification would naturally call for further work:

\begin{itemize}

\item It follows from the general results of Kozlov theory \cite{Koz1,Koz2,Koz3} (in
some cases, extended to include W-symmetries as well) that
equations of the form \eqref{eq:ffin} -- in some cases with extra
conditions on $S(t)$ -- can be integrated by passing to  symmetry
adapted variables. Thus each of our symmetric equations listed in Theorem 1 could be integrated by using the corresponding symmetries.

\item An alternative -- albeit strongly related -- approach to
integration of stochastic equations is that based on stochastic
invariants \cite{Koz18a,Koz18b,KozB}; it would be interesting to
have a similar classification for invariants of scalar stochastic
equations with environmental noise.

\item We have classified equations with \emph{environmental}
noise; it would be natural to consider the same task for equations
with \emph{demographic} noise, or equations with both
environmental and demographic noise (the so called \emph{complete
models} \cite{OvaMee}).

\end{itemize}

\noindent
These aspects lay outside our present scope and were not discussed here; we hope to tackle them in future work.

%\end{document}

\newpage

\centerline{{\Large{\bf SUPPLEMENTARY MATERIAL}}}

\bigskip\bigskip

This file provides supplementary material for the paper ``Symmetry
classification of scalar Ito equations with multiplicative
noise'', referred afterwards as ``the main paper''. In particular
we detail the computations needed to prove the second part of
Theorem 1 in there; in the following when we refer to Theorem 1 we
mean Theorem 1 in that paper.

Here we examine all the different possibilities (with some
duplication; i.e. the same situation is met under different
headings, see also Table A below) to have a solution to the
determining equations
\begin{eqnarray}
\vphi_t &+& f \, \vphi_x \ - \ \vphi \, f_x \ + \
\frac12 \Lap (\vphi ) \ = \ 0 \label{eq:deq1} \\
\vphi_w &+& \s \, \vphi_x \ - \ \vphi \, \s_x \ = \ R \, \s  \ ,
\label{eq:deq2} \end{eqnarray} where $\Lap$ is the Ito Laplacian,
which in our scalar case reads simply \beq \Lap ( \Psi ) \ := \
\frac{\pa^2 \Psi}{\pa w^2} \ + \ 2 \, \s \, \frac{\pa^2 \Psi}{\pa
x \pa w} \ + \ \s^2 \, \frac{\pa^2 \Psi}{\pa x^2} \ . \eeq We
identify all of these possibilities with one of the cases listed
in Theorem 1.

We have already constrained the forms of $f$ and $\varphi$ to
those appearing in equations (22) and (23) in the main paper (we
will use this notation throughout). We now identify all the
possible solutions of \eqref{eq:deq1} and \eqref{eq:deq2} by
refining the time dependence of the free parameters in $f$ and
$\varphi$. The relevant point is that in the equations above all
dependencies on $x$ can now be made explicit and thus the
coefficients of different monomials in $\{z,x,\log x\}$ should
vanish separately.

The computation is organized according  to $\a (t)$; we first
consider the generic case, i.e. the case where $\a(t) \not= \{- 1
, 0 +1 \}$, and then consider also the different special cases
$\a(t) = \{- 1 , 0 +1 \}$. Note that, by Remark 8 of the main
paper, in the special cases we can always set $\eta (t) = 0$.

\begin{appendix}

\section{Generic case}
\label{sec:generic}

We first require the coefficient of the term $x^{1 +\a} \log (x)$ to vanish;  this means
\beql{eq:xlogxg} - \, 2 \ \exp[ \a(t) \, S(t) \, z ] \  \a (t) \, S(t) \ \eta (t) \ R \  = \ 0 \ .   \eeq
Recalling that $S(t) \not= 0$ (or we would have a deterministic rather than a stochastic equation), and that $\a (t) \not= 0$ (or we would be in one of the special cases), we are left with the options
\begin{itemize}
\item {\bf (G.a)} $R \ = \ 0$, $\eta (t) \not= 0$;
\item {\bf (G.b)} $\eta (t) \ = \ 0$, $R \not= 0$;
\item {\bf (G.ab)} $R \ = \ 0$, $\eta (t) \ = \ 0$.
\end{itemize}
Note that the case $R=0$ yields only standard symmetries, while
conversely W-symmetry will occur only in the case $(G.b)$.

\subsection{Case (G.a): $R=0$, $\eta \not= 0$.}

Let us first consider the case \beq R \ = \ 0 \ , \ \ \eta (t) \not= 0 \ .  \eeq Then
\eqref{eq:xlogxg} is satisfied, and when we look at the
coefficient of $x^{1+\a(t)}$ we get \beq - \, 2 \ \exp[ \a(t) \,
S(t) \, z ] \  \a (t) \, S(t) \ \eta (t) \ Q_3 (t) \ . \eeq For
this to vanish -- as neither one of $S(t)$, $\a (t)$ and $\eta (t)$ can be zero under the present assumptions -- we need that
\beql{eq:genQ3} Q_3 (t) \ = \ 0  \ , \eeq and we assume this.
The equation \eqref{eq:xlogxg} is now a quadratic expression in $x$ (also depending on $z$), call it
$$ \sum_{k=0}^2 \ E_k \ x^k \ = \ 0 \ , $$ and this requires the $E_k$ to vanish separately; by looking at coefficients of $x^2$ and of $x^0$, we get
\begin{eqnarray*}
E_0 &=& - \, 2 \ \a(t) \, \[ 1 - \a^2 (t) \] \, \[1 + \a(t)\] \ S(t) \ F_1 (t) \ Q_2 (t) \ , \\
E_2 &=& 2 \ \a(t) \, \[ 1 - \a^2 (t) \] \, \[1 - \a(t)\] \ S(t) \ F_3 (t) \ Q_2 (t) \ . \end{eqnarray*}
Recalling that $S(t) \not= 0$ and that we assumed $\a (t) \not= \{ - 1 , 0 +1\}$, we have two options, viz.
\begin{itemize}
\item {\bf (G.a.1)} $Q_2 (t) \ = \ 0 $;
\item {\bf (G.a.2)} $F_1 (t) \ = \ F_3 (t) \ = \ 0 $.
\end{itemize}

In case {\bf (G.a.1)}, the equation \eqref{eq:deq1} is now fully satisfied; but this also enforces $\vphi = 0$, i.e. we get no symmetry altogether.

We will then consider case {\bf (G.a.2)}, i.e. assume
\beq F_1 (t) \ = \ F_3 (t) \ = \ 0 \ ; \ \ Q_2 (t) \ \not= \ 0 \ . \eeq Now the equation \eqref{eq:deq1} is homogeneous of degree one in $x$, and linear in $z$.

The coefficient of the term $x z$ reads
$$ 2 \, \a (t) \, [ 1 - \a^2 (t)] \, S(t) \, Q_2 (t) \ \[ S(t) \, \a' (t) \ + \ S' (t) \, \a (t) \] \ . $$
As $Q_2 (t) \not= 0$, we require that
$$ S(t) \, \a' (t) \ + \ S' (t) \, \a (t) \ = \ 0 \ , $$
which yields immediately
\beql{eq:a1a} \a (t) \ = \ \frac{k_1}{S(t)}  \ . \eeq
We stress that we should exclude both $k_1 = 0$ and the case where $S(t)$ is constant, $S(t) = s_0 \not= 0$ and $s_0 = \pm k_1$, as these would lead to $\a (t) = \pm 1$.

In this way \eqref{eq:deq1} is reduced to
\begin{eqnarray*} & &  k_1 \, x \, \( k_1^2 - S^2(t) \) \ \[ k_1 \, S^3 (t) \, Q_2 (t) \ + \ 2 \, k_1 \, Q_2 (t) \, S' (t) \right. \\
& & \left. + \ 2 \, S(t) \, Q_2 (t) \, \( S' (t) - k_1 F_2 (t) \) \ - \ 2 \, S^2 (t) \, Q_2' (t) \] \ = \ 0 \ . \end{eqnarray*}
This can be solved for $F_2(t)$, yielding
\beql{eq:F2} F_2 (t) \ = \ \frac{S^2 (t)}{2} \ + \ \frac{S' (t)}{k_1} \ + \ \frac{S'(t)}{S(t)} \ - \ \frac{S(t)}{k_1} \ \frac{Q_2' (t)}{Q_2 (t)} \ . \eeq

In this case we obtain (recall $k_1 \not= 0$)
\begin{eqnarray}
f(x,t) &=& x \ \[ S^2 (t) \ \( \frac12 \ + \ \frac{x^{k_1/S(t)} \, \eta (t) }{k_1 \, (k_1^2 - S^2 (t))} \) \ + \ \frac{S' (t)}{k_1} \right. \nonumber \\
& & \left. \ + \ \frac{\log(x) \, S' (t)}{S(t)} \ - \ \frac{S(t) \, Q_2' (t)}{k_1 \, Q_2 (t)} \] \ . \label{eq:gfa2} \end{eqnarray}
This expression gets simplified for $S(t) = s_0 \not= \pm k_1$; in this case we get
\beql{eq:fa2c}
f(x,t) \ = \ x \ \[ s_0^2 (t) \ \( \frac12 \ + \ \frac{x^{k_1/s_0} \, \eta (t) }{k_1 \, (k_1^2 - s_0^2)} \) \ - \ \frac{s_0 \, Q_2' (t)}{k_1 \, Q_2 (t)} \] \ . \eeq

As for the symmetries, in this case we get
\beq q(t,z) \ = \ \exp[ - \, k_1 \ z ] \ Q_2 (t) \ ; \eeq
going back to the original variables and the full symmetry coefficient, we get  \beql{eq:gpa2} \vphi (x,t,w) \ = \ x^{[1 + k_1/S(t)]} \ e^{- k_1 w} \ Q_2 (t) \ . \eeq
This shows that in this case we get nontrivial symmetries; these are (random) standard ones, as (they depend on $w$ and) $R=0$.

This case corresponds to case {\tt (a)} in the list of Theorem 1.

\subsection{Case (G.b): $\eta=0$, $R \not= 0$.}

We will now look at case {\bf (G.b)}, i.e. assume
\beql{eq:eta0}  R \ \not= \ 0 \ , \ \ \ \eta (t) \ = \ 0 \ . \eeq

Substituting for \eqref{eq:eta0} in \eqref{eq:deq1} we obtain an expression quadratic in $x$ and also depending linearly on on $\log(x)$, $z$ and on $\exp[\a (t) S(t) z ]$. The coefficient of $\exp[\a(t) S(t) z] \log(x)$ is
$$ - \, 2 \, R \, \a(t) \, [1 - \a^2 (t)] \ S(t) \ \[ F_1 (t) \ - \ x^2 \, F_3 (t) \] \ . $$
If we exclude the special cases $\a (t) = \{ - 1, 0 , +1\}$, recall that $S(t) \not= 0$, and exclude also $R=0$, for this to vanish identically in $x$ we need again
\beq F_1 (t) \ = \ 0 \ = \ F_3 (t) \ , \eeq
which hence we assume. This eliminates any dependence on $\log (x)$

Next we look at the coefficient of $z$ in \eqref{eq:deq1}, which is now
$$ 2 \, x \, \a(t) \, [1 - \a^2 (t) ] \, S(t) \ Q_2 (t) \ \[ S(t) \, \a' (t) \ + \ S' (t) \, \a(t) \] \ . $$
Requiring this to vanish, we have two cases:

\begin{itemize}
\item {\bf (G.b.1)} $Q_2 (t) = 0$;
\item {\bf (G.b.2)} $\a(t) = k_1 /S(t)$ where $k_1 \not= 0$ is a constant.
\end{itemize}

These cases look identical to subcases (G.a.1) and (G.a.2) above, but now we mean that (G.b) is also satisfied, while in there it was assumed that (G.a) holds. Moreover, now we do not have \eqref{eq:genQ3}, so $Q_2$ can be zero without implying $q(t,z)=0$.

\subsubsection{Case (G.b.1)}

Case {\bf (G.b.1)} is rather simple: the equation \eqref{eq:deq1} is then reduced to
$$ - \exp[z \a S] \, x \, \a (1 - \a^2) \ \[ R S^3 + 2 R S F_2 - 2 R S' - 2 Q_3 S' + 2 S Q_3' \] \ = \ 0 \ . $$
We can drop the initial factors and consider just the term in square brackets; moreover we write
\beql{eq:Q2Q3R} Q_3 (t) \ = \ R \ \theta (t)  \ ; \eeq hence
\beq R \ \[ S^3 (t) \ + \ 2 \, \( S(t) \, F_2 (t) \ - \ S' (t) \) \ - \ 2 \ \( \theta (t) \, S' (t) \ + \ S(t) \theta' (t) \) \] \ = \ 0 \ . \eeq
This -- and hence the full equation \eqref{eq:deq1} -- is solved by
\beq F_2 (t) \ = \ - \ \frac{S^3 (t) \ - \ 2 \, \( 1 \, + \, \theta' (t) \) \, S' (t) \ + \ 2 \, S(t) \, \theta' (t) }{2 \ S(t) } \ . \eeq

In this way we obtained
\beql{eq:gfb1} f(x,t) \ = \ - x \ \frac{S^3 (t) \ - \ 2 \(\log(x) \ + \ \theta (t) \) \, S' (t) \ + \ 2 \, S(t) \, \theta' (t)}{2 \ S(t)} \ . \eeq
For $S(t) = s_0$ this reduces to
\beq f(x,t) \ = \ - \ \frac{x}{2} \ \[ s_0^2 \ + \ 2 \ \theta' (t) \] \ . \eeq
As for the symmetries, in this case we get
\beq q(t,z) \ = \ R \ \theta (t) \ , \eeq
and hence in the original variables
\beql{eq:gpb1} { \vphi (x,t,w) \ = \ R \ x \ \[ \log(x) \ + \ \theta (t) \] } \ . \eeq
This corresponds to case {\tt (e)} in the list of Theorem 1, with $\F (t) = 0$.

\subsubsection{Case (G.b.2)}

Let us now consider case {\bf (G.b.2)}. Now the equation \eqref{eq:deq1} is linear in $x$ and also depends on $\exp [k_1 z]$. Looking at the coefficient of $x \exp[k_1 z]$ and dropping a coefficient $k_1 (k_1^2 - S^2)/S^3$, and using again \eqref{eq:Q2Q3R}, we are left with the equation
$$ S' \ - \ 2 (1 + \theta) S' \ + \ 2 S \( F_2 + \theta' \) \ = \ 0 \ , $$
which is solved by
\beq F_2 (t) \ = \ \frac{- S^3 (t) \ + \ 2 \, S' (t) \ + \ 2 \, \theta (t)  \, S' (t) \ - \ 2 \, S(t) \, \theta' (t)}{2 \ S(t) } \ . \eeq
With this, \eqref{eq:deq1} becomes homogeneous of degree one in $x$; dropping an inessential factor $2 k_1 R x (k_1^2 - S^2(t))$, we are left with
$$ k_1 \, S^3 \, Q_2 \ - \ k_1 \, Q_2 \, \theta \, S' \ - \ S^2 \, Q_2' \ + \ S \, Q_2 \, \( S' \ + \ k_1 \, \theta' \) \ = \ 0 \ . $$
This, and hence the full equation \eqref{eq:deq1} -- is solved by
\begin{eqnarray} Q_2 (t) &=& k_2 \, \exp \[ \int_0^t \frac{k_1 S^3(y) \, + \, S(y)  S' (y) \, - \, k_1  \[ \theta (y)  S' (y) \, + \, \theta' (y)  S(y) \]}{S^2 (y)} \, d y \] \nonumber \\
& & \ := \ H(t) \ . \end{eqnarray}
Note that for $S(t) = s_0$ this reduces to
\beq H(t) \ = \ H_0 (t) \ := \ k_1 \ \[ s_0 \ + \ \frac{\theta' (t)}{s_0 } \] \ . \eeq

In this way we obtain
\beql{eq:gfb2} f(x,t) \ = \ - x \ \frac{S^3 (t) \ - \ 2 \, \[ \log(x) \, + \, \theta (t) \] \, S' (t) \ + \ 2 \, S(t) \, \theta' (t) }{2 \ S(t) } \ ; \eeq
for $S(t) = s_0$ this reduces to
\beq f(x,t) \ = \ - \, x \ \[ \frac{s_0^2}{2} \ + \ \theta' (t) \] \ . \eeq

As for the symmetries, writing $k_2 = R k_3$, we get \beq q(t,z) \
= \ R \ \[ k_3 \ \exp\[H(t) \ - \ k_1 \, z \] \ + \ \theta (t) \]
\ , \eeq and hence \beql{eq:gpb2} {\vphi(x,t,w) \ = \ R \, x \,
\[ k_3 \, \exp \[ H(t) - k_1 w \] \, x^{k_1 / S(t)} \ + \ \log (x)
\ + \ \theta (t)
\] } \ . \eeq This corresponds to case {\tt (e)} in the list of
Theorem 1.

\subsection{Case (G.ab): $R = 0 = \eta(t)$}

In our discussion we have not allowed, so far, to have both $R=0$ and $\eta(t) = 0$; this represents the intersection of the two cases considered above. When we set
$$ R \ = \ 0 \ , \ \ \ \eta (t) \ = \ 0 \ , $$
the equation \eqref{eq:deq1} is of course simpler than in previous cases. Looking at the coefficient of $x^2 \exp[ - \a(t) S(t) z]$ we get
$$ 2 \ \a(t) \ [1 - \a^2 (t)] \ S(t) \ F_3 (t) \ Q_3 (t) \ = \ 0 \ . $$
We have again two cases:
\begin{itemize}
\item {\bf (G.ab.1)} $F_3 (t) \ = \ 0$;
\item {\bf (G.ab.2)} $Q_3 (t) \ = \ 0$.
\end{itemize}

\subsubsection{Case (G.ab.1)}

Let us first consider the case {\bf (G.ab.1)}. Now the coefficient of $x \exp[ - \a(t) S(t) z]$ vanishes for
\beq Q_3 (t) \ = \ k_3 \ S(t) \ ; \eeq
imposing this and looking at the coefficient of $\exp[ - \a(t) S(t) z]$ we obtain that moreover
\beq F_1 (t) \ = \ 0 \ . \eeq

At this point the equation \eqref{eq:deq1} is homogeneous of degree one in $x$, and linear in $z$, Dropping an overall factor $x \a(t) [1 - \a^2 (t)]$ and looking at terms of degree one in $z$ we get
$$ - \, 2 \ S(t) \ Q_2 (t) \ \[ S(t) \, \a' (t) \ + \ S'(t) \, \a (t) \] \ = \ 0 \ . $$
This is solved either for $Q_2 (t) = 0$ or for $\a(t) = k_1/S(t)$.

\bigskip
\noindent{\it Case (G.ab.1.1).}
For $Q_2 (t) = 0$ the full equation is also solved. In this case we get
\beql{eq:gfab11} f(x,t) \ = \ x \, F_2 (t) \ + \ x \, \[ \log (x) - 1 \] \frac{S' (t)}{S(t)} \ , \eeq
which for $S(t) = s_0$ just gives
\beq f(x,t) \ = \ x \, F_2 (t)  \ . \eeq

As for symmetries, we get
\beq q(t,z) \ = \ k_3 \ S(t) \ , \eeq
and therefore
\beql{eq:gpab11} { \vphi (x,t,w) \ = \ k_3 \ x \ S(t) } \ . \eeq
This corresponds to case {\tt (c)} in Theorem 1.

\bigskip
\noindent{\it Case (G.ab.1.2).}
In the case $\a(t) = k_1/S(t)$, the equation \eqref{eq:deq1} is finally solved setting
$$ F_2 (t) \ = \ \frac{S^2(t)}{2} \ + \ \frac{S'(t)}{k_1} \ + \ \frac{S'(t)}{S(t)} \ - \ \frac{S(t) \ Q_2' (t)}{k_1 \ Q_2 (t)} \ . $$

In this case we get
\beql{eq:gfab12} f(x,t) \ = \ \[ S^2 (t) \ + \ \frac{2}{k_1} \, \( S' (t) \, - \, S(t) \frac{Q_2'(t)}{Q_2(t)} \) \ + \ 2 \, \frac{S'(t)}{S(t)} \, \log (x) \] \ \frac{x}{2} \ , \eeq
which for $S(t) = s_0$ reduces to
\beq f(x,t) \ = \ \[ s_0^2 \ - \ \frac{2 s_0}{k_1} \, \frac{Q_2'(t)}{Q_2(t)} \] \ \frac{x}{2} \ . \eeq
As for symmetries, we get
\beq q(t,z) \ = \ k_3 \, S(t) \ + \ e^{- k_1 z} \ Q_2 (t) \ , \eeq
and hence
\beql{eq:gpab12} { \vphi (x,t,w) \ = \ k_3 \, S(t) \, x \ + \ Q_2 (t) \, e^{- k_1 w} \, x^{[1 + k_1/S(t)]} } \ . \eeq
This corresponds to case {\tt (b)} in Theorem 1.

\subsubsection{Case (G.ab.2)}

Let us now consider the case {\bf (G.ab.2)}, i.e. $Q_3 (t) = 0$. We have to annihilate an expression which is linear in $z$. The coefficient of $z$ is just
$$ 2 \ x \, \a(t) \, [1 - \a^2 (t)] \, S(t) \ Q_2 (t) \ \[ S(t) \a' (t) \ + \ S' (t) \, \a (t) \] \ . $$
If we set $Q_2 (t) = 0$ the full equation is satisfied, but we get $\vphi (x,t,w) = 0$, i.e. no symmetry.

By setting $\a(t) = k_1 /S(t)$ we are reduced to consider an expression quadratic in $x$ and not depending on $z$. The coefficients of $x^2$ and $x^0$ vanish if we require either $Q_2 (t) = 0$ or $F_1 (t) = F_3 (t) = 0$. In the case $Q_2 (t) = 0$ the full equation is satisfied, but $\vphi (x,t,w) = 0$. So we set
$$ F_1 (t) \ = \ 0 \ = \ F_3 (t) \ . $$
In this way we are left with an equation, homogeneous of degree one in $x$, which is solved by
\beq F_2 (t) \ = \ \frac{S^2 (t)}{2} \ - \ \frac{S(t) \, Q_2' (t)}{k_1 \, Q_2 (t)} \ + \ \( \frac{1}{k_1} \ + \ \frac{1}{S(t)} \) \ S' (t) \ . \eeq
This yields in turn

\beql{eq:gfab2} f(x,t) \ = \ \frac12 \ x \ \[ S^2 (t) \ - \ \frac{2 \, S(t) \, Q_2' (t) }{k_1 \, Q_2 (t) } \ + \ \frac{2}{k_1} \, S' (t) \ + \ \frac{2 \log(x) S'(t)}{S(t)}  \] \ , \eeq
which for $S(t) = s_0$ reduces to
\beq f(x,t) \ = \ \frac12 \ s_0 \ x \ \[ s_0 \ - \ \frac{2}{k_1 } \ \frac{Q_2' (t)}{Q_2 (t)} \] \ . \eeq

As for symmetries, we have
\beq q(t,z) \ = \ \exp[ - k_1 \ z ] \ Q_2 (t) \ , \eeq
and hence
\beql{eq:gpab2} { \vphi (x,t,w) \ = \ x^{[1+k_1/S(t)]} \ \exp[ - k_1 w  ] \ Q_2 (t) } \ . \eeq
This corresponds to case {\tt (b)} in Theorem 1.

We have thus concluded the discussion for the general case. We will now pass to consider special cases.

\section{Special case P: $\a(t) = +1$}

For $\a (t) = 1$, the equation \eqref{eq:deq1} contains a factor
$\exp[ - S(t) z ] x^2 \log^2(x)$; the vanishing of its coefficient
corresponds to $$ - \, 2 \ R \ \eta (t) \ = \ 0 \ , $$ hence we
must have one of the cases {\bf (a)}, {\bf (b)}, {\bf (ab)}
considered in the generic setting. These will now be dubbed as
cases (P.a), (P.b) and (P.ab) respectively.

\subsection{Case (P.a)}

It follows from Remark 8 of the main paper that we can disregard
the case $\eta (t) \not= 0$; we will however perform computations
in this case as well,to show we obtain no possibility of a
symmetric equation.

Let us first set
$$ R \ = \ 0 \ , \ \ \eta (t) \ \not= \ 0 \ . $$
The coefficient of $\exp[- S(t) z] x^2 \log(x)$ is
\beql{eq:PR0} - 2 \ \eta (t) \ Q_3 (t) \ . \eeq Thus we require
$$ Q_3 (t) \ = \ 0 \ , $$
and we have an equation linear in $z$. The vanishing of the
coefficient of $xz$ in this requires
$$ 4 \ S(t) \ Q_2 (t) \ S' (t) \ = \ 0 \ . $$
Choosing $Q_2 (t) = 0$ satisfies the full equation but produces
$\vphi=0$. Choosing instead
$$ S(t) \ = \ s_0 \ \not= \ 0 $$
gives an equation quadratic in $x$. For this to hold -- not allowing $Q_2 (t)
= 0$, which would give no nontrivial symmetry -- we would have to set
$$ \eta(t) \ = \ 0 \ , \ \ \ F_1 (t) \ = \ 0 \ , \ \ F_2 (t) \ = \
\frac{s_0^2}{2} \ - \ \frac{Q_2' (t)}{Q_2 (t)} \ . $$
But this is in contradiction with our assumption $\eta (t) \not= 0$. Thus in this case {\bf (P.a)} we have no nontrivial symmetries.

\subsection{Case (P.b)}

Let us consider case {\bf (P.b)}, i.e. the case where
\beq \eta (t) \ = \ 0 \ , \ \ R \ \not= \ 0 . \eeq
Looking at coefficients of terms involving $\exp[z S(t)] \log(x)$,
we see that we need
$$ F_1 (t) \ = \ 0 \ = \ F_3 (t) \ . $$
With this the equation is homogeneous of degree one in $x$, so we
drop the inessential overall $2 x$ factor. The coefficient of $z$
in the new equation is
$$ - \ 2 \ S(t) \ Q_2 (t) \ S' (t) \ , $$ so we have either $Q_2
(t) = 0$, which we call case{\bf (P.b1)}, or $S(t) = s_0$, which we call case {\bf (P.b2)}.

\bigskip\noindent
{\it Case (P.b1).}
For
$$ Q_2 (t) \ = \ 0 $$ it turns out we also have to require
\beql{eq:F2Pb1} F_2 (t) \ = \ - \ \frac{R \, S^3 (t) \ - \ 2 \[ R + Q_3 (t) \]
S' (t) \ + \ 2 \ S(t) \ Q_3' (t) }{2 \ R \ S(t) }\ . \eeq

In this way we obtain (once again $Q_3 = R \theta$)
\begin{eqnarray}
f(x,t) &=& - \( \frac{S^2 (t)}{2} \ - \ \[ \log (x) + \theta (t)
\] \frac{S' (t)}{S(t)} \ + \ \theta' (t) \) \ x \ , \label{eq:Pfb1} \\
\vphi(x,t,w) &=& R \ \[ \log (x) \ + \ \theta (t) \] \ x \ . \label{eq:Ppb1}
\end{eqnarray}
This corresponds to case {\tt (e)} of Theorem 1.

\bigskip\noindent
{\it Case (P.b2).}
If instead we set
$$ S(t) \ = \ s_0 \ \not= \ 0 \ , $$
then we also need (we write again $Q_3 = R \theta$)
$$ F_2 (t) \ = \ - \ \frac{s_0^2 \ + \ \theta' (t)}{2} $$
and moreover
$$ Q_2 (t) \ = \ k_1 \ \exp \[ s_0^2 t \ + \ \theta (t) \]
\ . $$

In this way we get (writing also $k_1 = R k_3$)
\begin{eqnarray}
f (x,t) &=& - \ \( \frac{s_0^2}{2} \ + \ \theta' (t) \) \ x
\label{eq:Pfb2} \\
\vphi(x,t,w) &=& R \ \( k_3 \ \exp \[ s_0^2 \, t \ - \ s_0 \, w \ + \
\theta (t) \] \ x^2 \ + \ x \, \log(x) \ + \ \theta (t)
\, x \) \ . \label{eq:Ppb2}  \end{eqnarray}
This corresponds to case {\tt (h)} of Theorem 1.

\subsection{Case (P.ab)}

If we set
$$ R \ = \ 0 \ , \ \ \eta (t) \ = \ 0 \ , $$
we obtain an equation which does not depend on $\log(x)$,
but still has $\exp[S(t) z]$ factors. In particular, looking at
coefficients of $\exp[S(t) z] x^2$ and $\exp[S(t) z]$ we must
require
\begin{eqnarray*}
4 \ S(t) \ F_1 (t) \ Q_3 (t) &=& 0 \ , \\
4 \ S(t) \ F_3 (t) \ Q_3 (t) &=& 0 \ ; \end{eqnarray*}
thus we have the two cases
\begin{itemize}
\item {\bf (P.ab.1)} $F_1 (t) = F_3 (t) = 0$,
\item {\bf (P.ab.2)}
$Q_3 = 0$.
\end{itemize}

\subsubsection{Case (P.ab.1)}

Setting
$$ F_1 (t) \ = \ 0 \ = \ F_3 (t)  $$ and looking at the coefficient of $\exp[S(t) z] x$ we obtain
$$ 4 \ \[ S(t) \, Q_3' (t) \ - \ S'(t) \, Q_3 (t) \] \ = \ 0 \ , $$
which requires
$$ Q_3 (t) \ = \ k_1 \ S(t) \ ; $$
note that here $k_1 = 0$ is a legitimate choice.

With this, \eqref{eq:deq1} is homogeneous of degree one in $x$ and
linear in $z$. The coefficient of $x z$ is
$$ - \, 4 \ S(t) \ Q_2 (t) \ S' (t) \ , $$
hence either $Q_2 (t) = 0$ or $S(t) = s_0$ is a constant.

\bigskip\noindent
{\it Case (P.ab.1.1).} For
$$ Q_2 (t) \ = \ 0 $$
the full equation is solved; now $F_2 (t)$ is unconstrained, and we get
\begin{eqnarray} f(x,t) &=& \[ F_2 (t) \ - \ \frac{S'(t)}{S(t)} \] \ x \ + \ \( \frac{S'(t)}{S(t)} \) \ x \ \log x \ , \label{eq:Pfab11} \\
\vphi(x,t,w) &=& k_1 \ S(t) \ x \ . \label{eq:Ppab11} \end{eqnarray}
This corresponds to case {\tt (c)} of Theorem 1.

\bigskip\noindent
{\it Case (P.ab.1.2).} For $S(t) = s_0$ the equation is reduced to
$$ 2 \, x \ \[ s_0^3 \, Q_2 (t) \ - \ 2 \, s_0 \ \( F_2 (t) \, Q_2 (t) \ + \ Q_2' (t) \) \] \ = \ 0 \ , $$
which is solved by
$$ F_2 (t) \ = \ \frac{s_0^2}{2} \ - \ \frac{Q_2' (t) }{Q_2 (t)} \ . $$

In this way we obtain \begin{eqnarray} f(x,t) &=& \( \frac{s_0^2}{2} \ - \
\frac{Q_2' (t)}{Q_2 (t)} \) \ x \ , \label{eq:Pfab12} \\
\vphi (x,t,z) &=& e^{- s_0 w} \ Q_2 (t) \ x^2  \ + \ k_1 \,
s_0 \, x \ . \label{eq:Ppab12} \end{eqnarray}
This corresponds to case {\tt (f)} of Theorem 1.

\subsubsection{Case (P.ab.2)}

If after choosing $R=0$ and $\eta(t)=0$ we do not set $F_1 = F_3 =
0$ but instead $$ Q_3 (t) \ = \ 0 \ , $$ we are left with an
equation linear in $z$. The coefficient of $x z$ is just
$$ - \, 4 \ S(t) \ Q_2 (t) \ S' (t) \ . $$ so either $S(t)$ is a
constant or
$$ Q_2 (t) \ = \ 0 \ . $$
Choosing this the equation is satisfied, but we get $\vphi (x,t,w)
= 0$, hence no symmetry.

Choosing instead
$$ S(t) \ = \ s_0 \ \not= \ 0 \ , $$
we are left with the two equations
\begin{eqnarray*}
8 \ s_0 \ F_1 (t) \ Q_2 (t) &=& 0 \ , \\
4 \, s_0 \ Q_2' (t) \ - \ 2 \, [ s_0^3 - 2 s_0 F_2 (t) ] \, Q_2
(t) &=& 0 \ . \end{eqnarray*} The case $Q_2 (t)=0$ was just
considered, so we exclude this and consider instead the other
option, i.e. \beq F_1 (t) \ = \ 0 \ , \ \ \ F_2 (t) \ = \
\frac{s_0^2}{2} \ - \ \frac{Q_2' (t)}{Q_2 (t)} \ . \eeq

In this way we obtain
\begin{eqnarray}
f(x,t) &=& \( \frac{s_0^2}{2} \, - \, \frac{Q_2'(t)}{Q_2 (t)} \) \
x \ + \ F_3 (t) \ x^2 \ , \label{eq:Pfab2} \\
\vphi (x,t,w) &=& Q_2 (t) \ \exp [ - s_0 w ] \ x^2 \ . \label{eq:Ppab2}
\end{eqnarray}
This corresponds to case {\tt (f)} of Theorem 1.

This concludes the discussion for the special case P.

\medskip\noindent
{\bf Remark.} In a recent work (see ref.[1] of the main paper) the
symmetries of the stochastic logistic equation
$$ d x \ = \ \( \a \, x \ - \ \b \, x^2 \) \ d t \ + \ \ga \ d w $$
(with $\a ,\b , \ga$ positive real constants) have been determined
and used to explicitly integrate it. In particular, it turned out
that in this case
$$ \vphi \ = \ \exp \[ - \( (\a - \ga^2/2) \, t \ + \ \ga \, w \) \] \ x^2 . $$
This fits within the case last considered, (P.ab.2), with the
choice $Q_2 (t) = 1$, $F_3 (t) = - \b$; note that here $s_0 =
\ga$.  \EOR

\section{Special case M: $\a (t) = - 1$}

When we set \beq \a(t) \ = \ - 1 \ , \eeq and look for the
vanishing of the coefficient of $\log^2 (x)$ in \eqref{eq:deq1},
we get \beq R \ \eta (t) \ = \ 0 \ . \eeq We have of course two
cases: {\bf (M.a)} $R=0$, and {\bf (M.b)} $\eta(t) = 0$; we will
consider the intersection of these two as a separate case {\bf
(M.ab)} for ease of discussion.

\subsection{Case (M.a): $R=0$, $\eta (t) \not= 0$.}

Looking at the coefficient of $\log (x)$, we have
$$ \eta (t) \ Q_3 (t) \ = \ 0 \ , $$
as we do not allow at present $\eta (t) = 0$, we set
$$ Q_3 (t) \ = \ 0 \ . $$
The whole equation has now a factor $\exp[z S(t)]$, which we drop from now on.

We are left with an expression quadratic in $x$; the term independent of $x$ is just
$$ - \ \eta (t) \ Q_2 (t) \ , $$
hence we set
$$ Q_2 (t) \ = \ 0 \ . $$
The equation is now satisfied but $\vphi = 0$, so we have no symmetry.

\subsection{Case (M.b): $\eta (t) = 0$, $R \not= 0$.}

We now consider the case with $R=0$ and $\eta(t) \not= 0$. The
coefficients of terms with $\log(x)$ enforce
$$ F_1 (t) \ = \ 0 \ = \ F_3 (t) \ . $$
In this way we have an expression homogeneous of degree one in $x$
(we drop from now on the $x$ factor), linear in $z$ and also
depending on $\exp[z S(t)]$.

The coefficient of $z \exp[z S(t)]$ is just
$$ 2 \ S(t)  \ Q_2 (t) \, S' (t) \ , $$ so either $Q_2 (t) = 0$ (case (M.b.1))
or $S(t) = s_0 \not= 0$ (case (M.b.2)).

\bigskip\noindent
{\it Case (M.b.1).} Setting
$$ Q_2 (t) \ = \ 0 $$ we are left with an equation which is solved by
$$ F_2 (t) \ = \ \frac{S'(t)}{S(t)} \ - \ \frac12 \, S(t) \ + \ \frac{Q_3 (t)}{R} \, \frac{S'(t)}{S(t)} \ - \ \frac{Q_3' (t)}{R} \ . $$

We thus obtain (writing as usual $ Q_3 = R \theta$)
\begin{eqnarray}
f(x,t) &=& x \, \log(x) \ \frac{S'(t)}{S(t)} \ + \ x \ \[ \theta(t) \, \frac{S'(t)}{S(t)} \ - \ \theta' (t) \ - \ \frac{S^2 (t)}{2} \] \label{eq:Mfb1} \\
\vphi (x,t,w) &=& R \ \( \log(x) \ + \ \theta (t) \) \ x \ . \label{eq:Mpb1} \end{eqnarray}
This corresponds to case {\tt (e)} of Theorem 1.

\bigskip\noindent
{\it Case (M.b.2).} If we set instead
$$ S(t) \ = \ s_0 \ \not= \ 0 \ , $$
the coefficient of the term $\exp[s_0 z]$ vanishes for
$$ F_2 (t) \ = \ \frac{s_0^2}{2} \ + \ \frac{Q_2' (t)}{Q_2 (t)} \ ; $$
the full equation is then solved (using again $Q_3 = R \theta$) by
$$ Q_2 (t) \ = \ k_1 \ \exp \[ - s_0^2 \, t \ - \ \theta (t) \] \ . $$

In this case we obtain
\begin{eqnarray}
f(x,t) &=& \[ \frac{s_0^2}{2} \ - \ \( s_0^2 \, + \, \theta' (t) \) \] \ x \label{eq:Mfb2} \\
\vphi (x,t,w) &=& R \ \( k_3 \, \exp \[ - s_0^2 t \ + \ s_0 w \ - \ \theta(t) \] \ + \ x \ \log (x) \ + \ \theta (t) \ x \) \ . \label{eq:Mpb2} \end{eqnarray}
This corresponds to case {\tt (h)} of Theorem 1.

\subsection{Case (M.ab): $R=0$, $\eta(t) = 0$.}

Finally we consider the subcase with
$$ R \ = \ 0 \ , \ \ \eta(t) \ = \ 0 \ . $$

The coefficient of $\exp[z S(t)] x^2$ is
$$ - \ 4 \ S(t) \ F_3 (t) \ Q_2 (t) \ . $$

\bigskip\noindent
{\it Case (M.ab.1).} Setting
$$ Q_2 (t) \ = \ 0 $$
and looking at the coefficient of $x$ we get
$$ Q_3 (t) \ = \ k_1 S(t) \ . $$
For $k_1 = 0$ the equation is solved but $\vphi=0$; so we assume
$k_1  \not= 0$. Then we also have to impose
$$ F_1 (t) \ = \ F_3 (t) \ = \ 0 \ . $$
In this way we get
\begin{eqnarray}
f(x,t) &=& \( F_2 (t) \ + \ \frac{S'(t)}{S(t)} \) \ x \ + \ \( \frac{S' (t)}{S(t)} \) \ x \ \log(x) \ , \label{eq:Mfab1} \\
\vphi(x,t,w) &=& k_1 \ S(t) \ x \ . \label{eq:Mpab1} \end{eqnarray}
This corresponds to case {\tt (c)} of Theorem 1.

\bigskip\noindent
{\it Case (M.ab.2).} Let us now consider instead the choice $F_3
(t) = 0$ (rather than $Q_2 (t) = 0$ as above). Then the
coefficient of $\exp[z S(t)] x z$ yields
$$ 2 \ S(t) \ Q_2 (t) \ S' (t) \ = \ 0 \ , $$
so we have the two options $Q_2 (t) = 0$ or $S(t) = s_0 \not= 0$.

\bigskip\noindent
{\it Case (M.ab.2.1).} Choosing $Q_2 (t) = 0$ we are left with the
two equations
$$ S(t) \ F_1 (t) \ Q_3 (t) \ = \ 0 \ , \ \ \ S(t) \, Q_3' (t) \ - \ S'(t) Q_3 (t) \ = \ 0 \ . $$
The case $Q_3 (t)= 0$ is of no interest here, (it yields $\vphi=0$) so we set
$$ F_1 (t) \ = \ 0 \ , \ \ \ Q_3 (t) \ = \ k_1 \ S(t) \ . $$
In this way we obtain
\begin{eqnarray}
f(x,t) &=& \( F_2 (t) \ + \ \frac{S' (t)}{S(t)} \) \ x \ + \ \frac{S' (t)}{S(t)} \ x \ \log (x) \ , \label{eq:Mfab21} \\
\vphi(x,t,w) &=& k_1 \ S(t) \ x \ . \label{eq:Mpab21} \end{eqnarray}
This corresponds to case {\tt (c)} of Theorem 1.

\bigskip\noindent
{\it Case (M.ab.2.2).} Finally, choosing $S(t) = s_0 \not= 0$
(instead of $Q_2 (t) = 0$) we get
$$ F_2 (t) \ = \ \frac{s_0^2}{2} \ + \ \frac{Q_2' (t)}{Q_2 (t)} \ ; $$ this in turn enforces
$$ F_1 (t) \ Q_3 (t) \ = \ 0 \ , \ \ Q_3' (t) \ = \ 0 \ . $$

For $Q_3 (t) = 0$ we get
\begin{eqnarray}
f(x,t) &=& F_1 (t) \ + \ \( \frac{s_0^2}{2} \ + \ \frac{Q_2'(t)}{Q_2(t)} \) \ x \ , \label{eq:Mfab22} \\
\vphi(x,t,w) &=& \exp [s_0 \ w ] \ Q_2 (t) \ . \label{eq:Mpab22} \end{eqnarray}
This corresponds to case {\tt (g)} of Theorem 1.

For $F_1 (t) = 0$, $Q_3 (t) = k_1$ we get instead
\begin{eqnarray}
f(x,t) &=& \( \frac{s_0^2}{2} \ + \ \frac{Q_2'(t)}{Q_2(t)} \) \ x \ , \label{eq:Mfab23} \\
\vphi(x,t,w) &=& k_1 \ x \ + \ \exp [s_0 \ w ] \ Q_2 (t) \ . \label{eq:Mpab23} \end{eqnarray}
This also corresponds to case {\tt (g)} of Theorem 1.

This concludes the discussion for the special case M.

\section{Special case O: $\a(t) = 0$}

The last special case we have to consider is that with
$$ \a (t) \ = \ 0 \ . $$
Setting this, and looking at the coefficients of terms with $\log
(x)$, we get that either $R=0$ or $F_1 (t) = F_3 (t) = 0$; we
denote these as cases (O.a) and (O.b) respectively. We will also
consider case (O.ab) with $R=F_1(t)=F_3 (t) = 0$.

\subsection{Case (O.a)}

Setting $R=0$ and looking at terms with $x^2$ and with $x^0$, we get
\begin{eqnarray*}
2 \ S(t) \ F_3 (t) \ \[ Q_2 (t) \ + \ Q_3 (t) \] &=& 0 \\
2 \ S(t) \ F_1 (t) \ \[ Q_2 (t) \ + \ Q_3 (t) \] &=& 0 \end{eqnarray*}

If we set $F_1 (t) = F_3 (t) = 0$ we are actually in case (O.ab);
so we disregard this option (for the time being) and set instead
$$ Q_3 (t) \ = \ - \ Q_2 (t) \ . $$
This fully satisfies equation \eqref{eq:deq1}; but it also yields
$\vphi = 0$, so in case {\bf (O.a)} we have no symmetry.

\subsection{Case (O.b)}

In case (O.b), in order to satisfy the equation it suffices to
choose $F_{2} (t)$ appropriately; this also fixes $f(x,t)$.
Writing now
$$ Q(t) \ = \ Q_2 (t) \ + \ Q_3 (t) \ = \ R \ \theta (t) \ , $$
we get
\begin{eqnarray}
f(x,t) &=& - \frac{x}{2 \, S(t)} \ \[ S^3 (t) \ - \ 2  S' (t) \( \log(x) + \theta (t) \) \ + \ 2 S(t) \theta' (t) \] \label{eq:Ofb} \\
\vphi (x,t,w) &=& R \ \( \log (x) \ + \ \theta (t) \) \ x \ . \label{eq:Opb} \end{eqnarray}
This corresponds to case {\tt (e)} of Theorem 1.

\subsection{Case (O.ab)}

Finally we consider the case with
$$ R=0 \ , \ \ F_1 (t) \ = \ F_3 (t) \ = \ 0 \ . $$
This is solved by choosing
$$ \eta (t) \ = \ \frac{\theta (t) \ S' (t) \ - \ \theta' (t) \ S(t)}{\theta (t) } $$
and in this way we get
\begin{eqnarray}
f(x,t) &=& \( F_2 (t) \ + \ [\log(x) \ - \ 1 ] \
\frac{\theta' (t)}{\theta (t)} \) \ x \label{eq:Ofab} \\
\vphi(x,t,w) &=& \theta (t) \ x \ . \label{eq:Opab} \end{eqnarray}
This corresponds to case {\tt (d)} of Theorem 1.

This concludes the discussion for the special case O, and hence all of our discussion.

\section{Conclusion}

We have exhausted the full classification of equations in the form
of eq.(3) in the main paper admitting random or simple
W-symmetries, i.e. equations described by vector fields of the
form (4) in the main paper. This was obtained by solving the
determining equations \eqref{eq:deq1}, \eqref{eq:deq2}.

The result of our detailed discussion is summarized in the Tables
provided in Theorem 1. It is maybe worth, for the ease of the
reader, to summarize the correspondence between the cases {\tt
(a)} -- {\tt (h)} considered in those Tables and in the statement
of the Theorem 1, and the cases considered for the sake of our
classification in this Supplementary Material. This is done in the
following Table A. \bigskip

\begin{tabular}{|c|l|}
\hline
  % after \\: \hline or \cline{col1-col2} \cline{col3-col4} ...
  cases in Theorem 1 & cases in this supplementary material \\
  \hline
  (a) & (G.a.2) \\
  (b) & (G.ab.1.2), (G.ab.2) \\
  (c) & (G.ab.1.1), (P.ab.1.1), (M.ab.1), (M.ab.2.1) \\
  (d) & (O.ab) \\
  (e) & (G.b.1), (G.b.2), (P.b1), (M.b.1), (O.b) \\
  (f) & (P.ab.1.2), (P.ab.2) \\
  (g) & (M.ab.2.2) \\
  (h) & (P.b2), M.b.2) \\
  \hline
\end{tabular}
\medskip

\noindent {{\tt Table A.} Correspondence between cases of Theorem
1 of the main text and cases considered in the present
Supplementary Material.}

\end{appendix}

\end{document}